\documentclass[11pt]{article}

\usepackage{graphicx,epsfig}
\usepackage{multicol}
\usepackage{amsmath,amssymb,cite,color,hyperref}
\newcommand{\be}{\begin{equation}}
\newcommand{\ee}{\end{equation}}
\newcommand{\bea}{\begin{eqnarray}}
\newcommand{\eea}{\end{eqnarray}}
\newcommand{\bi}{\begin{itemize}}
\newcommand{\ei}{\end{itemize}}
\newcommand{\ben}{\begin{enumerate}}
\newcommand{\een}{\end{enumerate}}

\newcommand{\lc}{\left[}
\newcommand{\rc}{\right]}
\newcommand{\lp}{\left(}
\newcommand{\rp}{\right)}
\newcommand{\as}{\alpha_s}

\def\frac#1#2{{{#1}\over {#2}}}
\def\gsim{\mathrel{\rlap{\lower4pt\hbox{\hskip1pt$\sim$}}
    \raise1pt\hbox{$>$}}}         
\def\lsim{\mathrel{\rlap{\lower4pt\hbox{\hskip1pt$\sim$}}
    \raise1pt\hbox{$<$}}}         

\newcommand{\draft}[1]{}

\definecolor{grey}{rgb}{0.5,0.5,0.5}

\begin{document}
\begin{flushright}
CERN-PH-TH/2011-225\\Edinburgh 2011/29\\FR-PHENO-2011-018\\IFUM-985-FT\\TTK-11-50
\end{flushright}
\begin{center}
{\large\bf Precision NNLO determination of
 $\alpha_s\lp M_Z\rp$  \\using an unbiased global parton set}
\vspace{0.6cm}

{\bf  The NNPDF Collaboration:}
Richard~D.~Ball$^{1}$, Valerio~Bertone$^2$,
 Luigi~Del~Debbio$^1$,\\ Stefano~Forte$^{3,4}$, Alberto~Guffanti$^5$, 
Jos\'e~I.~Latorre$^6$,  Simone~Lionetti$^{3}$, Juan~Rojo$^7$ and Maria~Ubiali$^{8}$.

\vspace{1.cm}
{\it ~$^1$ Tait Institute, University of Edinburgh,\\
JCMB, KB, Mayfield Rd, Edinburgh EH9 3JZ, Scotland\\
~$^2$  Physikalisches Institut, Albert-Ludwigs-Universit\"at Freiburg,\\ 
Hermann-Herder-Stra\ss e 3, D-79104 Freiburg i. B., Germany  \\
 ~$^3$ Dipartimento di Fisica, Universit\`a di Milano and\\
~$^4$ INFN, Sezione di Milano,\\ Via Celoria 16, I-20133 Milano,
Italy\\
~$^5$ The Niels Bohr International Academy and Discovery Center,\\
The Niels Bohr Institute, Blegdamsvej 17, DK-2100 Copenhagen, Denmark\\
~$^6$ Departament d'Estructura i Constituents de la Mat\`eria, 
Universitat de Barcelona,\\ Diagonal 647, E-08028 Barcelona, Spain\\
~$^7$ PH Department, TH Unit, CERN, CH-1211 Geneva 23, Switzerland \\
~$^8$ Institut f\"ur Theoretische Teilchenphysik und Kosmologie, RWTH Aachen University,\\ 
D-52056 Aachen, Germany\\}
\end{center}

\vspace{0.5cm}

\begin{center}
{\bf \large Abstract:}
\end{center}
We determine the strong  coupling $\alpha_s$ at NNLO in perturbative 
QCD using 
the global dataset input to the NNPDF2.1 NNLO parton fit: data from  
neutral and charged current deep-inelastic scattering, Drell-Yan, vector boson
production and inclusive jets. We find
$\alpha_s\lp M_Z\rp=0.1173\pm 0.0007^{\rm stat}$, where the statistical 
uncertainty comes from the underlying data and uncertainties due to
the analysis procedure are negligible.
We show that the distribution of $\alpha_s$ values preferred by
different experiments in the global fit is statistically
consistent, without need for rescaling uncertainties by a
``tolerance'' factor. We
show that if deep-inelastic data only are used, the best-fit value of
$\alpha_s$ is somewhat lower, but  consistent within one sigma with
the global determination. We find that the shift between the NLO and
NNLO values of $\alpha_s$ is $\Delta\alpha_s^{\rm pert}=0.0018$, and
we estimate the 
uncertainty  from higher-order corrections to be $\Delta\alpha_s^{\rm
  NNLO}\sim 0.0009$.  
\clearpage

We have recently shown~\cite{Lionetti:2011pw} that a surprisingly
precise determination of $\alpha_s$ using next-to-leading order 
perturbative QCD can be obtained from the
dependence on  the value of the strong coupling of the quality of a
global unbiased fit~\cite{Ball:2011uy} of parton distributions to deep-inelastic, Drell-Yan and
inclusive jet data. This result was somewhat unexpected. Indeed,
previous determinations of $\alpha_s$ from global parton
fits based on the idea of treating $\alpha_s$ as a fit 
parameter~\cite{Martin:2009bu,Alekhin:2009ni,Lai:2010nw} obtained 
larger uncertainties, and the  
precision of Ref.~~\cite{Lionetti:2011pw} is achieved even though the underlying
PDF
parametrization~\cite{Ball:2011uy} adopted there
is so general that the concept of a global best fit is no longer  
useful, and one must instead consider the fit
quality (as measured by the $\chi^2$) as a Monte Carlo random variable
which only 
stabilizes in the  limit of very large Monte Carlo sample size.

Given the great interest of an accurate determination of $\alpha_s$,
in particular for Higgs searches at the LHC~\cite{Dittmaier:2011ti},
it is important to repeat the analysis of Ref.~\cite{Lionetti:2011pw}
at NNLO, i.e. using the most accurate available QCD theory. This is
the purpose of this study: besides determining $\alpha_s$ and
providing a detailed assessment of  statistical and procedural
uncertainties, and a general discussion of its reliability, we will provide
an estimate of residual theoretical uncertainties.

Our methodology is the same as in
Ref.~\cite{Lionetti:2011pw}, to which we refer for a more detailed
discussion. Here, it will be sufficient to recall that our
determination is based on exploiting the fact that parton
distributions determined with the NNPDF methodology~\cite{DelDebbio:2004qj,DelDebbio:2007ee,
Ball:2008by,Ball:2009mk,Ball:2010de,Ball:2011gg} are delivered for
several  values of $\alpha_s$, with all other aspects of the analysis
being kept fixed. The quality of the fit of the data, as measured by
its $\chi^2$, can then be determined for each value of $\alpha_s$.
Because NNPDF parton distributions are delivered as a set of Monte
Carlo replicas, the $\chi^2$ is a random variable, whose fluctuations
only tend to zero in the limit of large sample size (as explicitly
verified in Ref.~\cite{Lionetti:2011pw}). The
expected standard deviation of the $\chi^2$ for given sample size can
be determined  using the  bootstrap method, as discussed in
Ref.~\cite{Lionetti:2011pw}. We thus arrive at a determination of the
$\chi^2$ curve as a function of $\alpha_s$ for a discrete set of
values of $\alpha_s$ and with an accuracy which depends on the size of
the Monte Carlo replica sample. Both the set of values
of $\alpha_s$ and the sample size can then be enlarged, until
satisfactory accuracy is reached.

\begin{table}[b!]
\begin{center}
\begin{tabular}{|c|c|}
\hline
 Values of $\alpha_s\lp M_Z\rp$ & $N_{\rm rep}$ \\
\hline
 0.114,0.115,0.116,0.117,0.118,0.119& 1000\\
 0.112,0.113,0.120,0.121 & 500 \\
 0.110,0.111,0.122,0.123,0.124 & 100 \\
\hline
\end{tabular}
\end{center}
\caption{\small The number of PDF replicas $N_{\rm rep}$ in the PDF
set for each value of $\alpha_s\lp M_Z\rp$. \label{tab:asnrep}}
\end{table}
In Ref.~\cite{Lionetti:2011pw} this was done by suitably enlarging
both the number of values of $\alpha_s$ and number of
replicas of the available  NNPDF2.1 NLO parton distribution
sets~\cite{Ball:2011mu}. Here, we start with the NNPDF2.1
NNLO parton distributions~\cite{Ball:2011uy}, which are  available
for $\alpha_s\lp M_Z\rp$ in the range from 0.114 to 0.124 in
steps of 0.001, with $N_{\rm rep}=1000$ replicas for $\alpha_s\lp
M_Z\rp=0.119$, and
$N_{\rm rep}=100$ replicas for all other values.  We have extended this range
down to 0.110, and we have considerably increased
the number of available replicas: we eventually end up
with the number of replicas and values of $\alpha_s$ summarized in
Table~\ref{tab:asnrep}.

\begin{figure}[t]
\centering
\epsfig{width=0.8\textwidth,figure=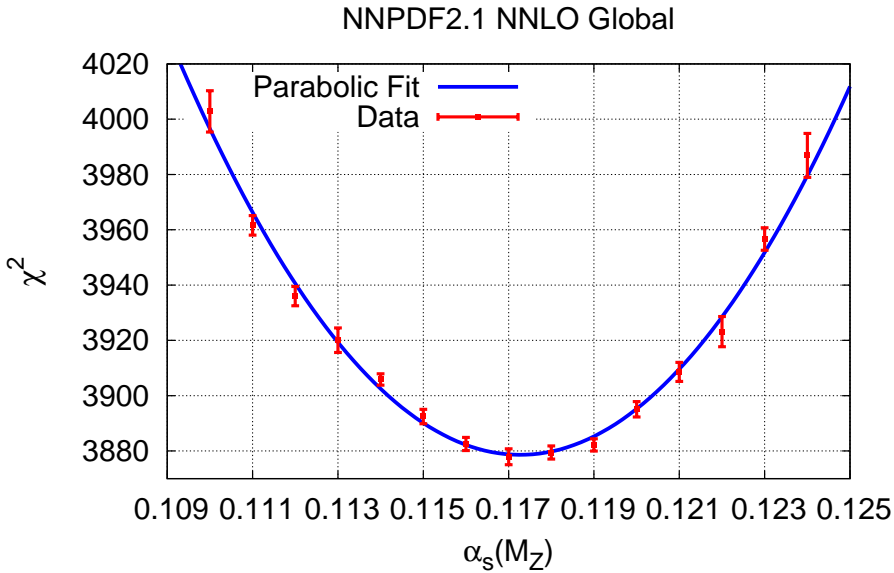}
\caption{\small The $\chi^2$ as a function of
$\alpha_s$ using  NNLO NNPDF2.1 parton distributions. The uncertainty
  shown is due to the finite size of the PDF replica sample used to
  compute the $\chi^2$ for each value of $\alpha_s$. The curve is the
  result of a parabolic fit.
\label{fig:chi2-profile}}
\end{figure}
The  $\chi^2$ values obtained for each value of $\alpha_s$ 
are shown in 
Fig.~\ref{fig:chi2-profile}; the total number of data points is
$N_{\rm dat}= 3357$~\cite{Ball:2011uy}, and at the minimum $\chi^2/N_{\rm dat}=1.16$.
The error bar shown on each point is
the standard deviation determined using the bootstrap method as
described in Ref.~\cite{Lionetti:2011pw}. A parabolic fit leads to
\be
\alpha_s^{\rm NNLO}\lp M_Z\rp=0.1173\pm 0.0007^{\rm stat} \pm
0.0001^{\rm proc} ,\label{eq:finres}
\ee
where the statistical uncertainty corresponds to a shift by one 
from the minimum of the parabola, while the procedural uncertainty is
propagated from the finite-size uncertainty on the $\chi^2$ values,
and it is clearly completely negligible. 
The $\chi^2$ per degree of freedom
of the parabolic fit shown in Fig.~\ref{fig:chi2-profile} is
$\chi^2_{\rm par}/N_{\rm dof}=1.1$. This value remains unchanged 
if the outer two or four points are removed from the parabolic fit:
this means that the behaviour of the
curve shown is parabolic to within the accuracy of the individual
points. 

Equation~(\ref{eq:finres}) is the main result of this
paper. It is interesting to compare this to the NLO result of
Ref.~\cite{Lionetti:2011pw}, namely
\be
\alpha_s^{\rm NLO}\lp M_Z\rp=0.1191\pm 0.0006^{\rm stat} \pm
0.0001^{\rm proc} ,\label{eq:nlores}
\ee
which was obtained using exactly the same methodology, from PDFs in
turn determined using the same methodology and the same
data~\footnote{The number of datapoints for the NLO fit of
  Ref.~\cite{Lionetti:2011pw} is  $N_{\rm dat}= 3338$~\cite{Ball:2011mu}: the slightly
  lower number of datapoints in the NLO case is due to the fact that
  the kinematic cuts on charm structure function data must be more
  stringent at NLO in order to exclude data for which NNLO corrections
  are very large, as discussed in more detail in Ref.~\cite{Ball:2011uy}.};
the value of the $\chi^2$ per data point  at the minimum at NLO is
also the same, namely $\chi^2/N_{\rm dat}=1.16$.
 The
result appears to be perturbatively quite stable. The small
difference in statistical error between the NLO and NNLO fits
indicates that the NLO fit quality deteriorates slightly faster when
moving away from the minimum. However, this difference 
does not appear to be statistically
significant: the gaussian uncertainty on the uncertainty is 
$\sigma_\sigma=\sigma/\sqrt{2 N}$, which in our
case, with $\sigma=0.0007$ and $N=N_{\rm dof}$ (the number of degrees
of freedom of the parabolic fit) is $\sigma_\sigma=0.0001$.

The statistical uncertainty on the determination of
$\alpha_s$ Eq.~(\ref{eq:finres}) is very small when compared to the
uncertainty on  other
existing determinations of $\alpha_s$: for example, the statistical 
uncertainty on
the NNLO MSTW determination~\cite{Martin:2009bu}, which is also based
on a PDF fit to a similar set of data,  is twice as large.
The size of the uncertainty on the MSTW08 result  is significantly
affected by the fact that a
tolerance~\cite{Pumplin:2001ct}  criterion is used to obtain 68\% confidence
levels. This means that  uncertainty ranges
in parameter space are rescaled by an amount which is determined by
requiring a reasonable behaviour of the 
distribution of best-fit results for the various
parameters among the different experiments that enter the global fit,
whose fluctuations may be significantly larger than expected on the
basis of unrescaled uncertainties. It is thus important to
establish whether such a rescaling is also necessary in our case.

\begin{figure}[ht!]
\begin{center}
\epsfig{file=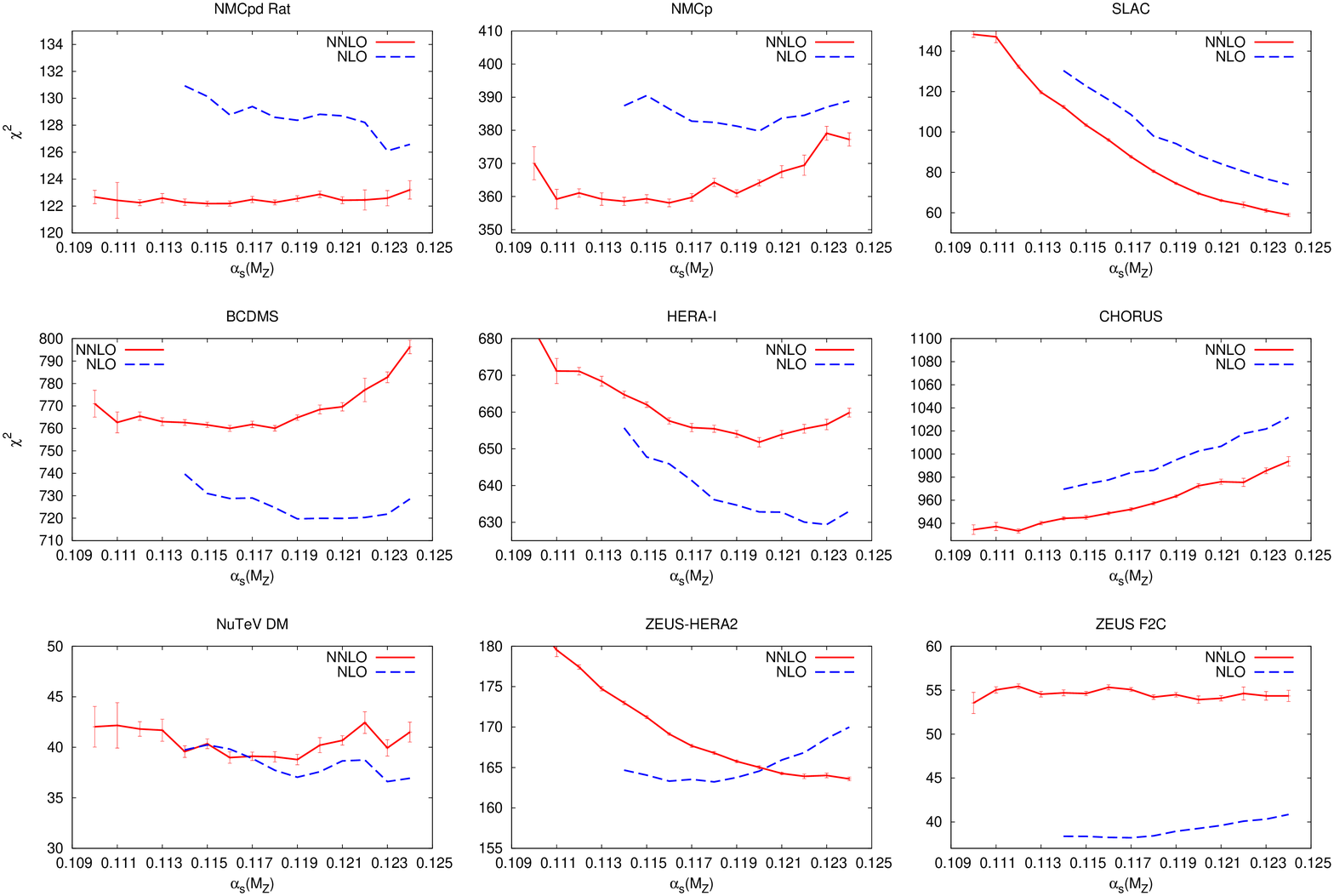,width=.97\textwidth}
\epsfig{file=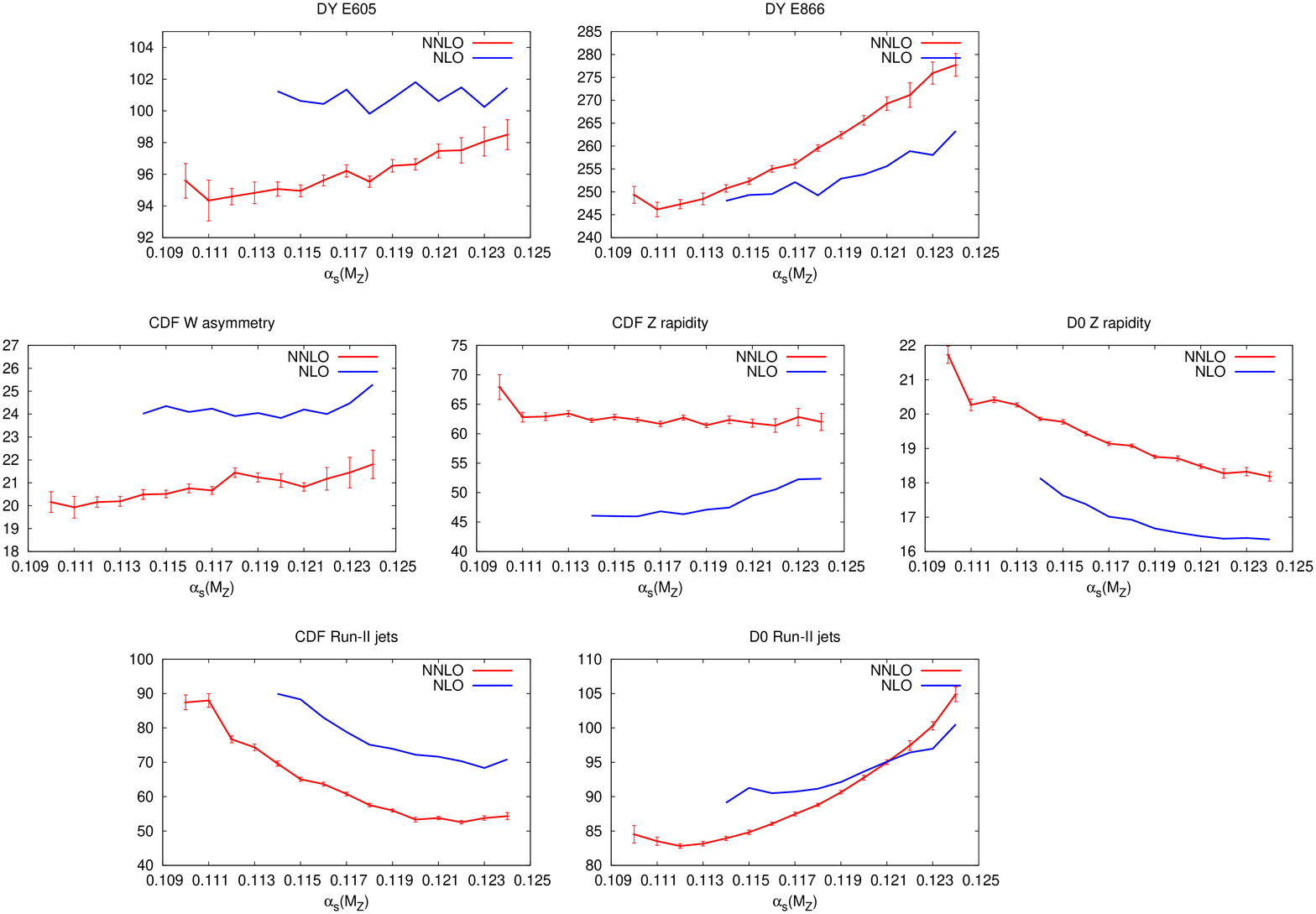,width=.97\textwidth}
\end{center}
\caption{\small Contribution to the total $\chi^2$ from
  each experiment entering the NNPDF2.1 NNLO PDF determination, plotted
  as a function of $\alpha_s$, including the uncertainty on each point
  due to the finite size of the replica sample. The NNPDF2.1 NLO curves from
  Ref.~\cite{Lionetti:2011pw} are also shown.
\label{fig:alphasfit-exps}}
\end{figure}

We do this (as in Ref.\cite{Lionetti:2011pw}) using a variant of the method
suggested in Ref.~\cite{Pumplin:2009sc}. Namely, we 
determine that the distribution of results obtained from individual
datasets  follows a
Gaussian distribution. A  possible deviation from unity of the width of this
distribution would then indicate the need for tolerance rescaling. Of
course  care should be taken that $\alpha_s$ values obtained
from individual experiments in a global fit are correlated by the
underlying global fit; these correlations could 
in principle lead to spurious minima 
in the $\chi^2$ profile which would be absent if $\alpha_s$ was fitted
from that experiment only, and thus would be of no
significance.


\begin{table}
\begin{center}
\begin{tabular}{|c|c|c|c|c|c|c|}
\hline
\hline
Experiment &     $\alpha_s $  &  $\sigma^{\rm stat}$  &  $\sigma^{\rm proc}$  &
 $\sigma^{\rm tot}$  &          $P$  & $\chi^{2}/N_{\rm dof}$   \\
\hline
\hline
NMCpd  &   0.1146 & 0.0105 & 0.0019 & 0.0107 & -0.25  & 0.5\\
NMCp  & 0.1150 & 0.0020 & 0.0003 & 0.0020 & -1.06 & 1.6 \\
BCDMS & 0.1158  & 0.0014 & 0.0002 & 0.0015 & -0.92 & 1.2 \\
HERA-I & 0.1199 & 0.0019  & 0.0002 & 0.0019 & 1.31 & 1.0  \\
ZEUS-H2 & 0.1231  & 0.0030 & 0.0002 & 0.0030 & 1.89 & 0.9 \\
NuTeV & 0.1177 & 0.0038 & 0.0004 & 0.0039 & 0.10 & 1.2 \\
CDFZRAP & 0.1205 & 0.0074 & 0.0033& 0.0081& 0.40 & 1.2 \\
CDFR2KT & 0.1225  & 0.0021 & 0.0003& 0.0021 & 2.35 & 1.8 \\
D0R2CON & 0.1111 & 0.0028& 0.0004& 0.0029 & -2.10 & 0.5 \\
\hline
\end{tabular}
\end{center}
\caption{\small The best-fit values of $\alpha_s$ 
obtained from a parabolic fit to the
  $\chi^2$ profiles of Fig.~\ref{fig:alphasfit-exps}. Only experiments
  for which the parabola has a minimum in the fitted range are
  included. In each case we also show the statistical and procedural
  uncertainties (defined as in Eq.~(\ref{eq:finres})), their sum in
  quadrature   $\sigma^{\rm tot}$, the 
pull $P$ Eq.~(\ref{eq:pull}), and the $\chi^2$ of the parabolic fit.
\label{tab:as-pulls}}
\end{table}

We have thus determined the profile of the contribution to the
 $\chi^2$ of the global fit from each experiment entering  the global
NNPDF2.1 NNLO determination: all profiles are  displayed in 
Fig.~\ref{fig:alphasfit-exps}, along with their NLO
counterparts of Ref.~\cite{Lionetti:2011pw}.  We have then determined
$\alpha_s$ and its  uncertainty from a parabolic fit to each
profile. For each experiment whose parabola has a minimum in the
fitted range we have determined the pull, defined as
\be
\label{eq:pull}
P_i\equiv \frac{\alpha_{s}^{ (i)}(M_Z)-\alpha_{s}(M_Z)}{\sqrt{\left(\sigma_{(i)}^{\rm tot}\right)^2+
\left(\sigma^{\rm tot}\right)^2}} \, ,
\ee
where $\alpha_{s}(M_Z)$ is the global best-fit value
Eq.~(\ref{eq:finres}) and 
$\sigma^{\rm tot}=\sqrt{\left(\sigma^{\rm stat}\right)^2+
\left(\sigma^{\rm proc}\right)^2}$ the uncertainty on it determined
from the statistical and procedural uncertainties also of
Eq.~(\ref{eq:finres}), while $\alpha_{s}^{(i)}(M_Z)$ and
$\sigma_{(i)}^{\rm tot}$ are the corresponding values for the $i$-th experiment.
 The best-fit value, statistical and procedural uncertainty for the
$i$-th experiment are determined by applying to the $\chi^2$ profile
for the contribution of the $i$-th experiment to the global fit
exactly the same procedure used for the determination of the
corresponding quantities in the global fit, and they are thus subject
to the caveat that correlations with other experiments are neglected
(for example, procedural uncertainties in each individual experiment
will now have point-by-point correlation due to other experiments
included in the global fit).

\begin{figure}[t]
\begin{center}
\epsfig{file=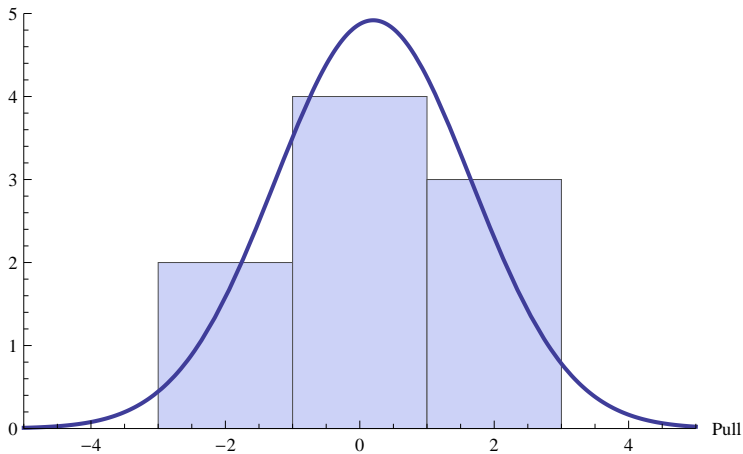,width=0.65\textwidth}
\caption{\small Histogram of the distribution of  pulls
included in Table~\ref{tab:as-pulls}, together with a gaussian fit to
it.
\label{fig:alphas-histo}}
\end{center}
\end{figure}
The best-fit values of $\alpha_s$, uncertainties and
pulls are collected in
Table~\ref{tab:as-pulls}, along with the $\chi^2$ of the parabolic fit
to each individual profile.\footnote{Note that the  definition of 
  pull Eq.~(\ref{eq:pull}) 
is the same as in Eq.~(3) of Ref.\cite{Lionetti:2011pw},
  however all pulls were incorrectly
given in Table~3 of that reference with the opposite
  sign convention.} Clearly  for some datasets (such as NMCpd or
CDFZRAP) the minimum is 
actually quite shallow, as apparent from 
Figure~\ref{fig:alphasfit-exps},  and the uncertainty in
Table~\ref{tab:as-pulls} is correspondingly large: these data would
not by themselves provide a determination of $\alpha_s$. However, note
that here we are not determining the best-fit $\alpha_s$ from each
dataset, but rather studying the compatibility of different data sets
within the global fit: from this point of view, data with a shallow
minimum or no minimum are certainly compatible, and thus not problematic.

The pull distribution has mean and standard deviation
\begin{equation}
\langle P\rangle=0.19\pm0.33;\quad \sigma_P=1.46\pm0.24,
\label{eq:pullfeat}
\end{equation}
where the uncertainties on the mean and standard deviation are
computed assuming that the population of nine pulls is extracted from a
univariate Gaussian distribution. 
In Fig.~\ref{fig:alphas-histo} 
the histogram of the pull
distribution is compared to a Gaussian normalized to the area under the histogram,
and with mean and width given by
Eq.~(\ref{eq:pullfeat}). 
We conclude that the pull distribution is consistent, with either no
need or minimal need for 
tolerance rescaling. Even if one were to rescale the statistical
uncertainty on $\alpha_s$ by the pull, one would end up with a statistical
uncertainty of 0.0010 instead of 0.0007.

Having established that the estimate of the statistical uncertainty in
our main result Eq.~(\ref{eq:finres}) is robust, we now turn
to the issue of theoretical uncertainties. The main theoretical
uncertainty comes from higher order QCD corrections. An indication of
its size can be obtained by comparing results obtained at subsequent
perturbative orders. Our NNLO and NLO results, Eqs.~(\ref{eq:finres})
  and~(\ref{eq:nlores}) respectively,  differ by
\begin{equation}
\Delta\alpha_s^{\rm pert}
\equiv \left|\alpha_s^{\rm NNLO}-\alpha_s^{\rm NLO}\right|= 0.0018.
\label{eq:nlonnlo}
\end{equation}
This is likely to provide an upper bound to the uncertainty on the NNLO
determination from  higher order corrections. Hence, the perturbative 
stability of our result suggests that the higher order uncertainty on
it is quite small.

A somewhat less crude estimate can be obtained using a method recently
proposed in Ref.~\cite{Cacciari:2011ze}, based on the idea that a
Bayesian estimate of the behaviour of unknown 
higher orders of a
perturbative series can be obtained from the behaviour of the known orders.
The 68\% confidence level (i.e. one sigma)
uncertainty $\Delta_k$ 
due to higher order corrections to the $k$-th order result
\be
\sigma_k = c_l\as^l+ c_{l+1}\as^{l+1}  + \dots + c_k\as^k \, 
\ee
under the assumptions of  
the model in  Ref.~\cite{Cacciari:2011ze} is then 
given by
\be
\Delta_k= \label{eq:cacciari} \left\{
\begin{array}{cc}
\as^{k+1}\max\{|c_l|,\dots, |c_k|\}\frac{n_c+1}{n_c}0.68 &	\mbox{ if }	0.68 \le \frac{n_c}{n_c+1}\\[10pt]
\as^{k+1} \max\{|c_l|,\dots, |c_k|\} \left[(n_c+1)0.32\right]^{-1/n_c}& \mbox{ if } 0.68 > \frac{n_c}{n_c+1}, 
\end{array}
\right. 
\ee
with $n_c = k+1-l$ the number of known perturbative coefficients. 

In our case, the output value of $\alpha_s$ in our analysis is 
not determined from the perturbative expansion of a single quantity, 
but rather from the expansion
for all processes which enter the global PDF fit. However, we can
obtain an estimate for its perturbative behaviour by letting
\be
\alpha_s^{\rm fit}\lp M_Z\rp=\sum_{k=0}^{\infty}c_k\lc \alpha_s^{0}
\lp M_Z\rp \rc^k
\label{eq:fakalphaexp}
\ee
where $\alpha_s^{\rm fit}$ is the result of our determination at a given
perturbative order, while $\alpha_s^{0}$ is some underlying input ``true''
value of the strong coupling.  
We then know the values of $\alpha_s^{\rm fit}$ at NLO and
NNLO, but we do not know the value of  $\alpha_s^{\rm fit}$ at LO, nor
the input value $\alpha_s^{0}$. We obtain our estimate by varying both
in a wide range: we take   $\alpha_s^{\rm fit,LO}\in \lc
0.100,0.180\rc$ and  $\alpha_s^{0}\in \lc0.110,0.125\rc$ and then we use
Eq.~(\ref{eq:cacciari}), with $\alpha_s^{\rm fit}$
Eq.~(\ref{eq:fakalphaexp}) evaluated at NLO or NNLO, and take in each
case the largest output value for $\Delta_k$ as the input parameters
are varied.

We then get
\bea
\Delta^{\rm NLO}\equiv\Delta_1=0.0077 \label{eq:nlothu}\\
\Delta^{\rm NNLO}\equiv\Delta_2=0.0009, \label{eq:nnlothu}
\eea
so that in particular the NNLO uncertainty is about half the shift
between the NLO and NNLO values. Clearly, results depend significantly
on the range of variation of $\alpha_s^{\rm fit,LO}$: for example, if
we restricted it to $\alpha_s^{\rm fit,LO}\in \lc
0.100,0.160\rc$ we would get $\Delta_1=0.0053$ and
$\Delta_2=0.0006$. We have chosen a very broad range in order to
obtain a conservative estimate.
A 
more careful assessment could be obtained by performing a
similar analysis (or factorization and renormalization scale
variation) for each of the processes which enters the global fit, and
then combining the ensuing uncertainties, while  the analysis given here
should be viewed as a semi-quantitative way of putting on a firmer
footing the expectation that the uncertainty on $\alpha_s$ 
at a given order due to higher-order corrections should be
a fraction of the shift of the best-fit value of $\alpha_s$ from the
previous to the given order.

We thus see that while at NLO theoretical uncertainties arising
from higher perturbative orders are dominant over
statistical uncertainties, 
 at NNLO the two uncertainties are of similar size, and thus our 
 determination of $\alpha_s$ appears to remain quite accurate even
 when this uncertainty is included.
Of the further sources of theoretical uncertainty, the largest  is
likely to be related to the treatment of heavy quarks, though such
uncertainty is typically~\cite{Ball:1995qd} 
smaller than that related to higher order corrections.

\begin{figure}[t]
\centering
\epsfig{width=0.8\textwidth,figure=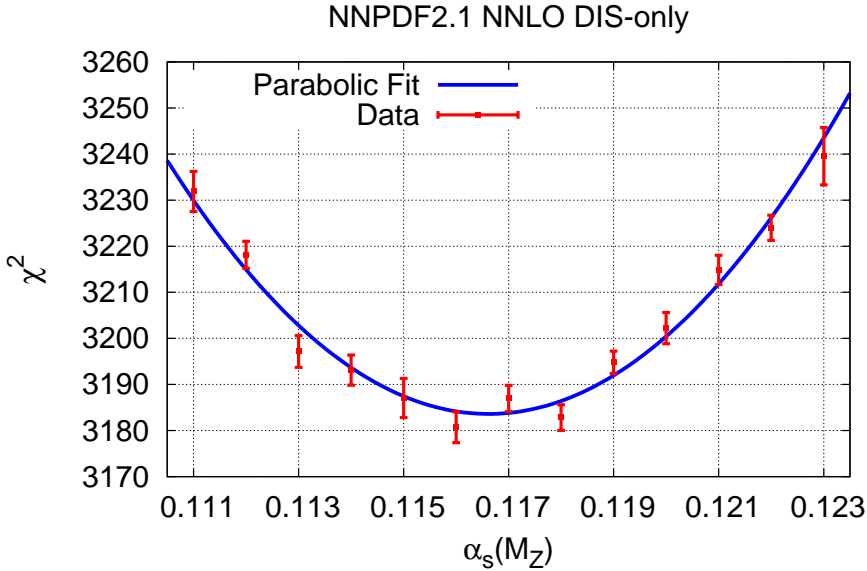}
\caption{\small Same as Fig.~\ref{fig:chi2-profile}, but using the
  NNPDF2.1 NNLO DIS only PDF set.
\label{fig:chi2-profile-dis}}
\end{figure}

A sizable fraction of the data which are used for the NNPDF2.1 PDF
determination comes from deep-inelastic scattering experiments. 
There is a widespread perception (see e.g.~\cite{Bethke:2009jm,Bethke:2011tr}) 
that $\alpha_s$ determinations based on deep-inelastic data lead
to a value which is somewhat smaller than the global average. In
Ref.~\cite{Lionetti:2011pw}, we have shown that in a DIS-only 
fit the BCDMS
data (which have been long
known~\cite{Virchaux:1991jc} to favor a low
$\alpha_s\sim0.114$)  have a minimum of the $\chi^2$ at rather low
$\alpha_s\lsim 0.113$. However, 
this is no longer the case if the BCDMS data are included in a global fit along
with Drell-Yan and jet data. We have traced this to the fact that
at low $\alpha_s$ DIS data and jet data pull the gluon in opposite
directions, so the fit quality can be improved using DIS data only in
a way which is forbidden when jet data are present.
Similar
results have been obtained in Refs.~\cite{Thorne:2011kq,munich}, where it was
further argued that this effect is stronger if the gluon
parametrization is too restrictive, and it was in fact shown
explicitly that within the MSTW fitting framework low 
NNLO best-fit values 
(similar or lower to the values $\alpha_s\sim0.113$ obtained at NNLO in
Ref.~\cite{Alekhin:2009ni}) are obtained if either the gluon
parametrization is more restrictive, or jet data are excluded from the
fit. In Ref.~\cite{Lionetti:2011pw} we found that in the
NNPDF framework (where an extremely flexible PDF parametrization is
used) the effect of the BCDMS data on the DIS fit remains moderate:
though the BCDMS data favor a low value of
$\alpha_s$, the DIS-only best fit $\alpha_s$ is still
$\alpha_s\sim0.118$, consistent with the global NLO best fit. 
However, the analysis of
Ref.~\cite{Lionetti:2011pw} is only performed at NLO, so one is
immediately led to ask what happens at NNLO.

\begin{figure}[h]
\begin{center}
\epsfig{file=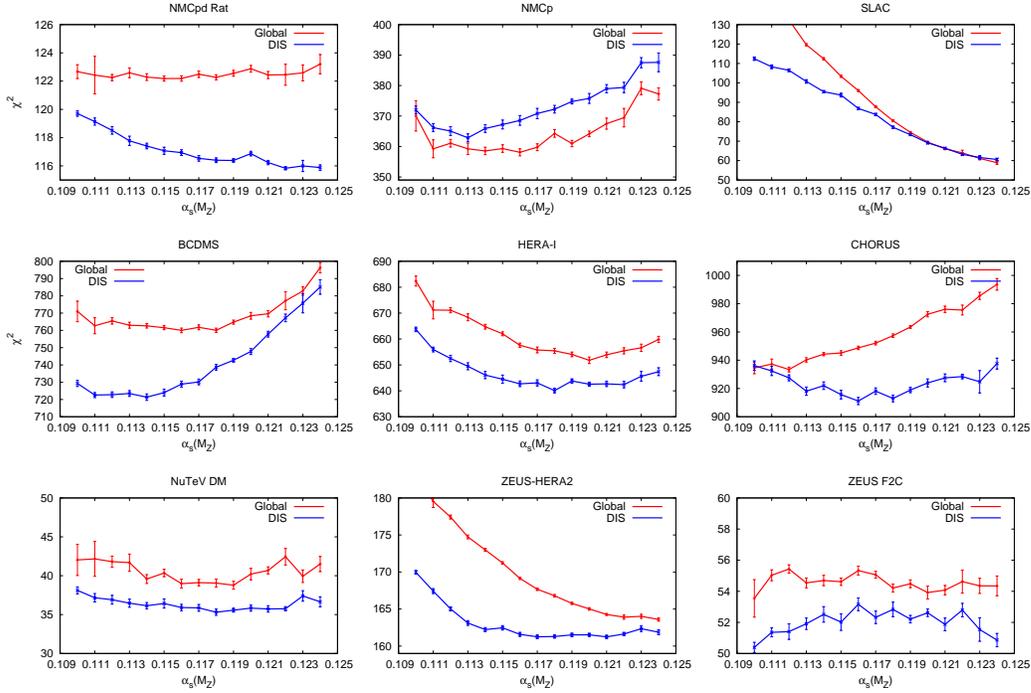,width=0.99\textwidth}
\end{center}
\caption{\small Same as Fig.~\ref{fig:alphasfit-exps}, but now
      comparing results obtained using the NNLO global and DIS-only fits.
\label{fig:alphasfitexps-dataset}}
\end{figure}

We have thus repeated our NNLO determination of $\alpha_s$, but now
only including deep-inelastic scattering data in the computation of
the $\chi^2$. Starting with the NNPDF2.1 NNLO DIS-only fit of
Ref.~\cite{Ball:2011uy}, we have generated $N_{\rm rep}=500$ replicas
for all values of $\alpha_s(M_Z)$ in the range from 0.110 to 0.124 in
steps of 0.001. The $\chi^2$ values and finite-size 
uncertainties that we get are shown in
Fig.~\ref{fig:chi2-profile-dis}, to be compared to
Fig.~\ref{fig:chi2-profile} of the global fit; the number of data
points now is $N_{\rm dat}=2783$~\cite{Ball:2011uy}, and at the minimum $\chi^2/N_{\rm
  dat}=1.14$.  A parabolic fit has a 
$\chi^2_{\rm par}/N_{\rm dof}=1.2$, and leads to
\be
\alpha_s^{\rm NNLO}\lp M_Z\rp=0.1166\pm 0.0008^{\rm stat} \pm
0.0001^{\rm proc} .\label{eq:finresdis}
\ee
Note that the outer two points  at $\alpha_s(M_Z)=0.110,\>0.124$
have not been used, because
 the quality of the parabolic fit would considerably deteriorate.
The $\chi^2$ profiles from individual datasets in the global and
DIS-only fit are compared in Fig.~\ref{fig:alphasfitexps-dataset}.

Comparing our global and DIS-only determinations of $\alpha_s$
Eqs.~(\ref{eq:finres}) and~(\ref{eq:finresdis}) we see that, like at
NLO~\cite{Lionetti:2011pw},  also at
NNLO the  DIS-only value, though somewhat smaller than the global one,
agrees with it within the (small) statistical uncertainty at the one
sigma level. The pattern of $\chi^2$ profiles for individual
experiments is also similar at NLO and NNLO: the BCDMS data have a
minimum for much lower $\alpha_s$ in the DIS only fit than in the
global fit.  Note however that
the overall DIS-only best-fit value of $\alpha_s$
Eq.~(\ref{eq:finresdis})
remains quite close
to the global one, unlike in the MSTW08 case where the NNLO DIS-only value
ends up being quite low (0.1139, to be compared to 0.1171 for the
global fit) --- perhaps 
because of our  especially flexible PDF parametrization. In this
respect, it is interesting to observe that MSTW
find~\cite{Thorne:2011kq} that their DIS-only result depends strongly
on whether the parametrization allows  the gluon to become negative or
not; in the NNPDF determination positivity is imposed at the level of
physical observables, thus leading to a milder constraint on the PDF itself.

\begin{figure}[t]
\begin{center}
\epsfig{file=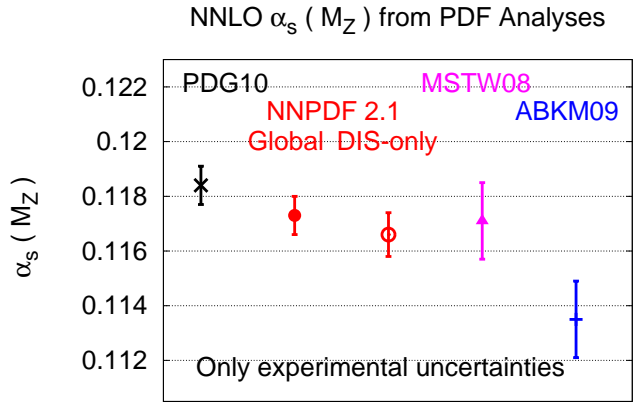,width=0.60\textwidth}
\end{center}
\caption{\small Comparison of the NNPDF2.1 NNLO
$\alpha_s\lp M_Z\rp$ determinations to other recent determinations
from NNLO PDF analysis. The PDG value of Ref.~\cite{Bethke:2009jm}
is also shown.
\label{fig:alphas-res-comp}}
\end{figure}

Our results for the NNLO $\alpha_s$ determinations, both from the
global data set and from DIS data only, are displayed graphically in 
Fig.~\ref{fig:alphas-res-comp}, where they are also compared
to the other recent NNLO 
ABKM09~\cite{Alekhin:2009ni} and MSTW08~\cite{Martin:2009bu}
determinations. Our results are in good agreement with MSTW, but with
smaller uncertainties. The ABKM result is rather lower, but it should be
kept in mind that ABKM only use DIS data and a smaller set of
fixed-target Drell-Yan data, with a parton parametrization which is
more restrictive than the MSTW one, so it is subject to the caveats
mentioned above and discussed in Ref.~~\cite{Thorne:2011kq}.

In summary, we have presented the most accurate available NNLO
$\alpha_s$ determination from a global parton fit. The greater statistical
accuracy of our result in comparison to other available determinations
of $\alpha_s$ along with PDFs~\cite{Martin:2009bu,Alekhin:2009ni} is
due both to the use of the particularly flexible and reliable NNPDF
approach to PDF parametrization and fitting, and to the use of a wide
and up-to-date dataset (in particular the recent combined HERA
data~\cite{h1:2009wt} which are not used in
Refs.~\cite{Martin:2009bu,Alekhin:2009ni}). The
theoretical
uncertainty on our determination is likely to be dominated by higher-order
perturbative QCD corrections, and thus there is no reason why it
should be different from any other analogous determination (such as
Refs.~\cite{Martin:2009bu,Alekhin:2009ni}). This uncertainty has not
been determined accurately so far. However,
the considerable perturbative stability we find, demonstrated by the
moderate shift between our NLO and NNLO results, suggests that it
might be smaller than hitherto expected: in particular, our estimate
for it is less than half than that provided in
Ref.~\cite{Martin:2009bu}. It will be interesting to repeat this
analysis once the very accurate data on standard candle  processes
from the LHC become available.

{\bf\noindent  Acknowledgments \\}
S.F. thanks A.~Martin for communicating Ref.~\cite{munich} prior to
publication, and G.~Altarelli and R.~Thorne for illuminating discussions.
J.~R. is supported by a Marie Curie 
Intra--European Fellowship of the European Community's 7th Framework 
Programme under contract number PIEF-GA-2010-272515.
M.U. is supported by the Bundesministerium f\"ur Bildung and Forschung (BmBF) of the Federal 
Republic of Germany (project code 05H09PAE).
This work was 
partly supported by the Spanish
MEC FIS2007-60350 grant. 
We would like to acknowledge the use of the computing resources provided 
by the Black Forest Grid Initiative in Freiburg and by the Edinburgh Compute 
and Data Facility (ECDF) (http://www.ecdf.ed.ac.uk/). The ECDF is partially 
supported by the eDIKT initiative (http://www.edikt.org.uk).
\bigskip


\end{document}